\documentstyle[preprint,aps,eqsecnum]{revtex}
\begin{document}
\vskip 1.4cm
\begin{center}
{\large\bf Solving the Darwin problem in the first post-Newtonian 
approximation of general relativity: compressible model}
\end{center}
\vspace{2mm}
\begin{center}
Masaru Shibata$^*$ and Keisuke Taniguchi$^{\dagger}$ \\
\vspace{5mm}
$^*${\small\em Department of Earth and Space Science, 
Graduate School of Science,
Osaka University,\\ Toyonaka, Osaka 560, Japan}\\
$^{\dagger}${\small\em 
Department of Physics, Kyoto University, Kyoto 606-01, Japan}
\end{center}
\begin{abstract}
Using the ellipsoidal model for the density configuration, 
we calculate the equilibrium sequence of the corotating 
binary stars of the polytropic equation of state 
in the first post-Newtonian approximation of 
general relativity. After we calibrate this model by comparing with 
previous numerical results, we perform the stability 
analysis by calculating the energy and the angular momentum of the 
system as a function of the orbital separation. We find that 
the orbital angular velocity at the energy and/or momentum minimum 
increases with the increase of the compactness of each star, and 
this fact holds irrespective of the polytropic index. 
These features agree with those in previous numerical works. 
We also show that due to the influence of the tidal field from the companion 
star, the central density of each star slightly decreases. 
\end{abstract}

{\small PACS number(s): 04.30.Db, 04.25.Nx, 04.40.Dg}

\break
\newcommand{\beq}{\begin{equation}}
\newcommand{\eeq}{\end{equation}}
\newcommand{\beqn}{\begin{eqnarray}}
\newcommand{\eeqn}{\end{eqnarray}}
\newcommand{\pa}{\partial}
\def\bI{\hbox{$\,I\!\!\!\!-$}}

\section{Introduction}

In a previous paper\cite{TS}, we analytically 
obtain the equilibrium configuration 
of corotating binary stars(i.e., the solution of 
the Darwin problem) of incompressible fluid in the first 
post-Newtonian(PN) approximation. We calculate the energy and 
angular momentum as a function of the orbital separation, and 
investigate the location of their minimums(we call the orbit at 
the minimum the innermost stable corotating circular 
orbit(ISCCO)\footnote{The ISCCO is the orbit where the secular instability 
sets in, and different from the innermost stable circular orbit(ISCO) 
where the dynamical instability sets in. The ISCO will locate 
inside the ISCCO\cite{LRS}.}). 
It is found that the angular velocity at the ISCCO increases with 
the increase of the compactness of the star; i.e., the general 
relativistic(GR) 
effect makes the angular velocity at the ISCCO larger. 

In this paper, we try to extend the previous work to the case of 
the compressible fluid star using the ellipsoidal model for the 
density profile(the so called ellipsoidal approximation) 
proposed by Lai, Rasio and Shapiro\cite{LRS}. 
The ellipsoidal approximation may be used for stars of 
the sufficiently stiff equation of state(EOS) of a large 
adiabatic index. Thus, this method may make it possible 
to investigate the ISCCO 
for realistic binary neutron stars which are made of the 
stiff EOS of the large adiabatic index $\Gamma \sim 2-3$\cite{shapiro}. 
In previous 
papers, one of us(M.S.)\cite{shibapn} numerically calculates the equilibrium 
state of binary neutron stars of the polytropic EOS with $\Gamma=2$ and 3 
in the PN approximation. We will show that 
by a simple semi-analytic calculation, the results 
obtained in previous papers can be qualitatively explained well. 
Also, we investigate the influence due to the 
GR tidal effect in binary neutron stars. 

In section II, we show the basic equations to derive the angular 
velocity for the corotating binary stars. In section III, we 
introduce the ellipsoidal approximation assuming the polytropic EOS. 
Using the ellipsoidal approximation, in section IV, we 
calculate the equations for the energy and angular momentum. 
In section V, we derive the scalar virial relation in the 
PN approximation, which is used to determine the size of the star. 
In section VI, we show equilibrium 
sequences of binary stars for various polytropic indices. 
Paying attention to the energy and angular momentum as a 
function of the orbital separation, we investigate the ISCCO. 
We also show that due to the influence of the PN tidal field from the companion
star, the central density of each star slightly decreases. 
Section VII is devoted to summary. 

Throughout this paper, we take the units of $G=1$, where $G$ is 
the gravitational constant, and use $c$ as the light velocity. 
Latin indices $i$, $j$, $k$, $l$, $\cdot \cdot$ take 
1 to 3, and $\delta_{ij}$ denotes the kronecker's delta. 
$M$ denotes the Newtonian mass for each star of the binary as
\beq
M=\int \rho d^3x. 
\eeq
The quadrupole moment and its tracefree part of each star are defined as 
\beq
I_{ii}=\int \rho x_i^2 d^3x,\hskip 5mm {\rm and} \hskip 5mm
\bI_{ii}=I_{ii}-{1 \over 3}\sum_l I_{ll}.
\eeq

\section{Formulation}

In this section, we show the basic equations deriving the equations of 
the angular velocity and the mass for corotating binary stars 
in the first PN approximation with no 
restriction to a special model because these quantities can be 
written in a simple form without specifying a density configuration. 
To construct the equilibrium sequence and argue the stability, 
we additionally 
need the equations for the energy, the angular momentum, and the 
scalar virial relation in the PN approximation. We derive them after 
we specify the model for the density configuration because 
they are not written in a simple form. 

We consider corotating binary stars of equal Newtonian 
masses($M_1=M_2=M$), whose coordinate separation is $R$, 
in the first PN approximation. 
Here, the gauge condition is the standard PN one\cite{chand}. 
We adopt the coordinates where the center mass of 
a star(star 1) locates at the origin and the other one(star 2) locates 
at $(x_1,x_2,x_3)=(-R,0,0)$. 
As for the density configuration, in this section, 
we only assume that the star has the triplane symmetry with respect 
to the symmetric plane(i.e., for star 1, $\rho(x,y,z)=\rho(-x,y,z)=
\rho(x,-y,z)=\rho(x,y,-z)$). In the following, when the integral symbol 
appears, it means that we perform the integral inside the star 1. 

In deriving the equations for various quantities, 
we assume $R > a_0 > M/c^2$, where $a_0$ is the 
typical size of the star: We calculate the terms up to $O((a_0/R)^{3})$ 
relative to the lowest order for the angular velocity and the 
angular momentum, and the terms to $O((a_0/R)^{4})$ for the energy. 
Then, the effects by the quadrupole moment of each star 
to the equilibrium state are taken into account consistently. 

We adopt the polytropic EOS as 
\beq
P=K\rho^{\Gamma}=(\Gamma-1)\rho\varepsilon,\hskip 1cm \Gamma=1+{1 \over n},
\eeq
where $P$, $\rho$, $\varepsilon$, $n$ and $K$ are the pressure, 
the mass density, the specific internal energy, the polytropic index and 
the polytropic constant, respectively. Then, 
in the first PN approximation, the integrated form of the Euler equation is 
written as\cite{PNeom}\cite{shibapn}\cite{asada}
\beqn
& &K'\rho^{\Gamma-1}-{1 \over 2c^2}\biggl(K'\rho^{\Gamma-1}\biggr)^2 
\nonumber \\
&=&U-{X_0 \over c^2}
+\biggl\{{\varpi^2 \over 2}+{1 \over c^2}\Bigl(2\varpi^2U-X_{\Omega}
+ \beta_{\varphi}\Bigr)\biggr\}\Omega^2
+{\varpi^4 \over 4c^2}\Omega^4+{\rm constant},\label{feq}
\eeqn
where $\varpi^2=(x_1+R/2)^2+x_2^2$ and $K'=K(n+1)$. 

The angular velocity is obtained from the first tensor virial(TV) relation
as 
\beqn
0=\int {\pa P \over \pa x_1}\biggl(1-{\Gamma \varepsilon \over c^2}
\biggr) d^3x 
&=&\int \rho U_{,1} d^3x+{R \over 2}M\Omega^2 
+{1 \over c^2}\biggl[-\int \rho X_{0,1} d^3x \nonumber \\
&& +\Omega^2_{\rm N}
\int\rho \Bigl(2\varpi^2U_{,1}+4x_1U +2RU-X_{\Omega,1}+ \beta_{\varphi,1}
\Bigr)d^3x \nonumber \\
&& ~~~
+\Omega^4_{\rm N} 
\biggl({R \over 2}(3I_{11}+I_{22})+{MR^3 \over 8}\biggr)\biggr]
,\label{tveq}
\eeqn
where $,k$ denotes the partial derivative with respect to $x_k$.
$\Omega_{\rm N}$ denotes the angular velocity in the Newtonian order 
solution as\cite{LRS}
\beq
\Omega_{\rm N}^2={2M \over R^3}+{18\bI_{11} \over R^5}. 
\eeq
To determine the orbital angular velocity up to the 
PN order, we need to perform the 
following integrals: 
\beqn
&&I_1=\int \rho U d^3x,\\
&&I_2=\int \rho U_{,1} d^3x ,\\ 
&&I_3=\int \rho U x_1 d^3x ,\\
&&I_4=\int \rho U_{,1} x_1 d^3x ,\\
&&I_5=\int \rho U_{,1} x_k^2d^3x,\\
&&I_6=\int \rho X_{0,1} d^3x,\\
&&I_7=\int \rho X_{\Omega,1} d^3x,\\
&&I_8=\int \rho \beta_{\varphi,1} d^3x. 
\eeqn
In the following subsections, we  separately show the equations for 
the gravitational potentials and their solutions to 
evaluate $I_1 \sim I_8$. 

\subsection{Equation and solution for $U$ and 
integrals including $U$: $I_1 \sim I_5$}
 
Equation for the Newtonian potential $U$ is 
\beq
\Delta U = -4\pi\rho. 
\eeq
Here, there are two kinds of contribution for $U$ at star 1; one is 
the contribution from star 1 itself and the other is that from star 2. 
Hence, we write $U$ as  
\beq
U=U^{1 \rightarrow 1} +  U^{2 \rightarrow 1} ,
\eeq
where $U^{i \rightarrow j}$ denotes the potential generated by star $i$ 
at star $j$. Here, $U^{2 \rightarrow 1}$ is generally expressed as 
\beqn
U^{2 \rightarrow 1}
&=&{M \over R}\biggl(1-{x_1 \over R}+{2x_1^2-x_2^2-x_3^2 \over 2R^2}
+{-2x_1^3+3x_1(x_2^2+x_3^2) \over 2R^3} +O(R^{-4})\biggr) \nonumber \\
& &+{3\bI_{11} \over 2 R^3}\biggl(1-{3x_1 \over R}+O(R^{-2}) \biggr). 
\eeqn
On the other hand, we cannot specify $U^{1 \rightarrow 1}$ 
without the density configuration. Thus, we 
define the following integrals\cite{Ch69}; 
\beqn
W&=&{1 \over 2}\int \rho U^{1 \rightarrow 1} d^3x,\\
{\cal M}_{ii}&=&\int \rho U^{1 \rightarrow 1}_{,i}x_i d^3x,
\eeqn
where we have the relation $\sum_i {\cal M}_{ii}=-W$\cite{Ch69}. 
Using the solution for $U$, we get 
\beqn
&&I_1=2W+{M^2 \over R}+{3M \over R^3}\bI_{11}+O(R^{-5}),\\
&&I_2 = -{M^2 \over R^2}-{9M\bI_{11} \over R^4}+O(R^{-6}),\\
&&I_3 = -{M \over R^2}I_{11}+O(R^{-4}),\\
&&I_4 ={\cal M}_{11}+{2M \over R^3}I_{11} + O(R^{-5}),\\
&&I_5=-{M \over R^2}I_{kk}+O(R^{-4}).
\eeqn
Note that $\Omega_{\rm N}$ is derived using $I_2$. 

\subsection{Equation and solution for 
$X_0$ and integrals including $X_0$: $I_6$}

As in the case of $U$, $X_0$ is divided into two parts as 
$X_0=X_0^{1 \rightarrow 1}+X_0^{2 \rightarrow 1}$. 
We also write $I_6$ as $I_6^{1 \rightarrow 1}+I_6^{2 \rightarrow 1}$, 
where $I_6^{1 \rightarrow 1}$ and $I_6^{2 \rightarrow 1}$ 
correspond to the contributions from 
$X_0^{1 \rightarrow 1}$ and $X_0^{2 \rightarrow 1}$, respectively. 
Equations for $X_0^{1 \rightarrow 1}$ and $X_0^{2 \rightarrow 1}$ are 
\beqn
& &\Delta X_0^{1 \rightarrow 1} 
= 4\pi\rho\biggl[2U^{1 \rightarrow 1}+2U^{2 \rightarrow 1}
+\varepsilon + {3P \over \rho} \biggr] \equiv 
4\pi\rho^{1 \rightarrow 1}_X, \\
& &\Delta X_0^{2 \rightarrow 1}
= 4\pi\rho\biggl[2U^{2 \rightarrow 2}+2U^{1 \rightarrow 2}
+\varepsilon + {3P \over \rho}\biggr]\equiv 4\pi\rho_X^{2 \rightarrow 1}. 
\eeqn
To evaluate $I_6^{1 \rightarrow 1}$, we need only the odd function of $x_1$ 
in $\rho^{1 \rightarrow 1}_X$, and the terms required are  
\beq
\rho^{1 \rightarrow 1}_X=-2\rho {x_1 \over R^2}\biggl(M+
M{2x_1^2-3x_2^2-3x_3^2 \over 2R^2}+{9 \bI_{11} \over 2R^2} 
+O(R^{-4})\biggr).
\eeq
Then, the terms of $X_0^{1 \rightarrow 1}$ 
required for $I_6^{1 \rightarrow 1}$ are 
\beq 
X_0^{1 \rightarrow 1}=\biggl({2 M \over R^2}+{9\bI_{11} \over R^4}\biggr)
D_1+{M \over R^4}(2D_{111}-3D_{122}-3D_{133})+O(R^{-6}),
\eeq
where $D_i$ and $D_{1kk}$ are the solutions of 
\beqn
& & \Delta D_i=-4\pi\rho x_i,\\
& & \Delta D_{1kk}=-4\pi\rho x_{1}x_k^2. 
\eeqn
Using the relations, 
\beqn
& &\int \rho D_{i,i} d^3x = -{\cal M}_{ii},\\
& &\int \rho D_{1kk,1} d^3x=-\int \rho U^{1 \rightarrow 1}_{,1}
x_1 x_k^2 d^3x \equiv 
-{\cal M}_{11kk}, 
\eeqn
$I_6^{1 \rightarrow 1}$ is evaluated to be 
\beq
I_6^{1 \rightarrow 1}= -\biggl({2 M \over R^2 }+{9\bI_{11} \over R^4}
\biggr){\cal M}_{11}-{M \over R^4}(2{\cal M}_{1111}-3{\cal M}_{1122}
-3{\cal M}_{1133})+O(R^{-6}). 
\eeq

On the other hand, the solution of $X_0^{2 \rightarrow 1}$ is written as 
\beqn
X_0^{2 \rightarrow 1}
&=&{M_X \over R}\biggl(1-{x_1 \over R}+{2x_1^2-x_2^2-x_3^2 \over 2R^2}
+{-2x_1^3+3x_1x_2^2+3x_1x_3^2 \over 2R^3}+O(R^{-4})\biggr) \nonumber \\
& &+{d_X \over R^2}\biggl(1-{2x_1 \over R}+O(R^{-2})\biggr)
+{3\bI_{X11} \over 2R^3}\biggl(1-{3x_1 \over R}+O(R^{-2})\biggr),
\eeqn
where
\beqn
&&M_X=-\int\rho_X^{2 \rightarrow 1} d^3x=-4W-(3\Gamma-2)U_{\rm i}
-2\biggl({M^2 \over R}+{3M\bI_{11} \over R^3}\biggr)+O(R^{-5}),\\
&&d_X=-\int\rho_X^{2 \rightarrow 1}x_1 d^3x=-{2MI_{11} \over R^2}
+O(R^{-4}),\\
&&\bI_{X11}=-{1 \over 3}\int\rho_X^{2 \rightarrow 1}(2x_1^2-x_2^2-x_3^2) d^3x
\nonumber \\
&& \hskip 1cm =-2\bI_{U11}-(3\Gamma-2)\bI_{\varepsilon 11}-{2M \over R}\bI_{11}
+O(R^{-3}),
\eeqn
and 
\beqn
&& U_{\rm i}=\int\rho\varepsilon d^3x,\\
&&\bI_{U11}={2I_{U11}-I_{U22}-I_{U33}  \over 3}, \hskip 1cm 
I_{Ukk}\equiv \int\rho U^{1\rightarrow 1}x_k^2 d^3x, \\
&&\bI_{\varepsilon 11}={2I_{\varepsilon 11}-I_{\varepsilon 22}-
I_{\varepsilon 33} \over 3}, \hskip 1.4cm 
I_{\varepsilon kk}\equiv \int\rho \varepsilon x_k^2 d^3x.
\eeqn
Thus, $I_6^{2 \rightarrow 1}$ is evaluated to be 
\beqn
I_6^{2 \rightarrow 1}&=&-M\biggl({M_X \over R^2}+{2d_X \over R^3}
+{9M_X\bI_{11} \over 2MR^4}+{9\bI_{X 11} \over 2R^4} \biggr)\nonumber \\
&=& M\biggl[ { 4W+(3\Gamma-2)U_{\rm i} \over R^2} + {2M^2 \over R^3}
+{ 24M\bI_{11} \over R^5 }+{4MI_{11} \over R^5}\nonumber \\
&&~~~+{9 \over 2R^4} \biggl\{ \Bigl( 4W+(3\Gamma-2)U_{\rm i} \Bigr)
{\bI_{11} \over M}+
2\bI_{U11}+(3\Gamma-2)\bI_{\varepsilon 11} 
\biggr\}+O(R^{-6})   \biggr].
\eeqn

\subsection{Equation and solution for $X_{\Omega}$ 
and integral including $X_{\Omega}$: $I_7$}

Equation for $X_{\Omega}$ is 
\beq
\Delta X_{\Omega} = 8\pi\rho\biggl(x_1^2+x_2^2+Rx_1+{R^2 \over 4}\biggr).
\eeq
The solution is also written as $X_{\Omega}^{1 \rightarrow 1}+
X_{\Omega}^{2 \rightarrow 1}$, where 
\beqn
X_{\Omega}^{1 \rightarrow 1}&=&
-2\biggl(D_{11}+D_{22}+RD_1+{R^2 \over 4}U^{1 \rightarrow 1}\biggr) , \\
X_{\Omega}^{2 \rightarrow 1}&=&
-{R^2 \over 2}U^{2 \rightarrow 1}+2{I_{11} \over R}
\biggl(1-{2x_1 \over R}+O(R^{-2})\biggr)
-{2(I_{11}+I_{22}) \over R}
\biggl(1-{x_1 \over R}+O(R^{-2})\biggr), \label{xom}
\eeqn
and $D_{kk}$ is the solution for $\Delta D_{kk}=-4\pi\rho x_k^2$. 
Using these solutions, we soon get $I_7$ as
\beq
I_7=2R{\cal M}_{11}+
{M \over 2}\biggl(M+{9\bI_{11} \over R^2}-{4 \over R^2}(I_{11}-I_{22})
+O(R^{-3})   \biggr).
\eeq

\subsection{Equation and solution for $ \beta_{\varphi}$ and 
integral including $\beta_{\varphi}$: $I_8$}

Definition of $\beta_{\varphi}$ is 
\beqn
\beta_{\varphi}&=&-{7 \over 2}\biggl(x_1 P_1+ x_2P_2+{R \over 2}P_1\biggr) 
\nonumber \\
& &-{1 \over 2}\biggl[\biggl(x_1+{R \over 2}\biggr)^2P_{2,2}+x_2^2P_{1,1}-
\biggl(x_1+{R \over 2}\biggr)x_2(P_{1,2}+P_{2,1})\biggr],
\eeqn
where 
\beqn
& &\Delta P_1 = -4\pi\rho \biggl(x_1+{R \over 2}\biggr), \\
& &\Delta P_2 = -4\pi\rho x_2.
\eeqn
$P_i$ is also written as $P_i^{1 \rightarrow 1}+P_i^{2 \rightarrow 1}$, 
where
\beqn
P_1^{1 \rightarrow 1} &=& D_1 +{R \over 2} U^{1 \rightarrow 1}, \\
P_2^{1 \rightarrow 1} &=& D_2,
\eeqn
and 
\beqn
P_1^{2 \rightarrow 1} &=& {I_{11} \over R^2}\biggl(1-{2x_1 \over R}
+{6x_1^2-3x_2^2-3x_3^2 \over 2R^2}+O(R^{-3})\biggr) 
- {R \over 2} U^{2 \rightarrow 1}, \\
P_2^{2 \rightarrow 1} &=& {I_{22} \over R^2}\biggl({x_2 \over R}
-{3x_1x_2 \over R^2}+O(R^{-3}) \biggr). 
\eeqn
$\beta_{\varphi}^{2 \rightarrow 1}$ can be written to the 
explicit form as 
\beqn
\beta_{\varphi}^{2 \rightarrow 1}&=&
{MR \over 8}\biggl(7+{7x_1 \over R}-{14x_1^2+9x_2^2+7x_3^2 \over 2R^2}
+{x_1(14x_1^2 +13x_2^2+7x_3^2) \over 2R^3}+O(R^{-4})\biggr) \nonumber \\
&+&{21\bI_{11} \over 16R}\biggl(1-{x_1 \over R}+O(R^{-2})\biggr)
-{7I_{11} \over 4R}\biggl(1+O(R^{-2})\biggr)
-{I_{22} \over 8R}\biggl(1+{x_1 \over R}+O(R^{-2})\biggr).
\eeqn 
Then, after straightforward calculations, $I_8$ is evaluated as 
\beq
I_8=-{R \over 2}(7W-{\cal M}_{22})
+{7M^2 \over 8}+{M \over 8R^2}(14I_{11}+9I_{22}+7I_{33})+O(R^{-3}).
\eeq

\subsection{$\Omega^2$}

Gathering the results obtained in previous subsections, $\Omega^2$ is 
calculated as 
\beqn
\Omega^2&=&{2M \over R^3}\biggl[1+{1 \over Mc^2}
\biggl\{ 4W+(3\Gamma-2)U_{\rm i}+{\cal M}_{33}-{\cal M}_{11}
-{9M^2 \over 4R}\nonumber \\
&&\hskip 2cm
-{M \over 2R^3}\Bigl(28I_{11}-16I_{22}-5I_{33}\Bigr)+O(R^{-4})\biggr\}\biggr]
\nonumber \\
&+&{18 \bI_{11} \over R^5} \biggl[ 1+{1 \over c^2M}\biggl(W
-{\cal M}_{11}-{\cal M}_{22}+{3\Gamma-2 \over 2}U_{\rm i} 
+O(R^{-2})\biggr)\biggr] \nonumber \\
&+&{1 \over c^2R^5}\biggl(18\bI_{U 11}+9(3\Gamma-2)\bI_{\varepsilon 11}
-4{\cal M}_{1111}+6{\cal M}_{1122}+6{\cal M}_{1133}
\biggr).
\eeqn
To transform this equation into a more simple form, we use the scalar, 
the second and the fourth tensor virial equations in the 
Newtonian order as follows;
\beqn
&& 3(\Gamma-1)U_{\rm i}=W-{3M \over R^3}\bI_{11}
-\Omega_{\rm N}^2(I_{11}+I_{22})
+O(R^{-5}),\\
&& {\cal M}_{11}=-(\Gamma-1)U_{\rm i}-\Omega_{\rm N}^2I_{11}
-{2MI_{11} \over R^3}+O(R^{-5}),\\
&& {\cal M}_{22}=-(\Gamma-1)U_{\rm i}-\Omega_{\rm N}^2I_{22}
+{MI_{22} \over R^3}+O(R^{-5}),\\ 
&& {\cal M}_{33}=-(\Gamma-1)U_{\rm i}+{MI_{33} \over R^3}+O(R^{-5}),\\
&& 2{\cal M}_{1111}-3{\cal M}_{1122}-3{\cal M}_{1133}
=-9(\Gamma-1)\bI_{\varepsilon 11}+
O(R^{-3}). 
\eeqn
Then, 
\beqn
\Omega^2&=&{2M \over R^3}\biggl[1+
{1 \over Mc^2}\biggl\{ 5W+U_{\rm i}-{9M^2 \over 4R}-
{M \over 2R^3}\Bigl(28I_{11}-14I_{22}-9I_{33}\Bigr)+O(R^{-4})\biggr\}\biggr]
\nonumber \\
&+&{18 \bI_{11} \over R^5} \biggl[1+{1 \over c^2}\biggl(
{13+n \over 6}{W \over M}+O(R^{-2})\biggr)\biggr] 
+{1 \over c^2R^5}\biggl(18\bI_{U11}+9(5\Gamma-4)\bI_{\varepsilon 11}
\biggr).
\eeqn
We note that in the incompressible case($\rho=$constant and 
$\Gamma\rightarrow \infty$), the configuration of each star 
is the ellipsoidal figure of its axial length $a_1$, $a_2$, and $a_3$, 
and we have relations
\beqn
& &W={2M\pi\rho A_0 \over 5},\\
& &U_{\rm i}=0,\\
& &\bI_{U11}={16\pi\rho A_0 \over 21}\bI_{11},\\
& &(\Gamma-1)\bI_{\varepsilon 11}={2\pi\rho A_0 \over 21}\bI_{11},
\eeqn
where\cite{Ch69}
\beqn
A_0&=&a_1a_2a_3 \int^{\infty}_0{du \over
\sqrt{(a_1^2+u)(a_2^2+u)(a_3^2+u)} } \nonumber \\
&=&a_1^2 \alpha_2\alpha_3 \int^{\infty}_0{dt \over
\sqrt{(1+t)(\alpha_2^2+t)(\alpha_3^2+t)} } \equiv a_1^2
\hat A_0(\alpha_2,\alpha_3),
\eeqn
and $\alpha_2=a_2/a_1$ and $\alpha_3=a_3/a_1$. Then, $\Omega^2$ 
reduces to that derived in a previous paper\cite{TS}. 

\subsection{Mass and center of mass}

The conserved mass is defined as\cite{chand}\cite{shibapn}
\beqn
M_* &=&\int d^3x \rho_* = \int d^3x \rho\biggl[1+{1 \over c^2}
\biggl( {v^2 \over 2}+3U \biggr)\biggr] \nonumber \\
& = &M\biggl[1+{1 \over c^2}\biggl( {6W \over M}+
{13M \over 4R}+{1 \over 4R^3}(45\bI_{11}+4I_{\rm s})
+O(R^{-5}) \biggr) \biggr],
\eeqn
where $I_{\rm s}=I_{11}+I_{22}$. 
The PPN mass\cite{will} is slightly different from $M_*$ as 
\beqn
M_{\rm PPN} &=& \int d^3x \rho\biggl[1+{1 \over c^2}
\biggl\{ {v^2 \over 2}+\varepsilon+3U-{U^{1\rightarrow 1} \over 2}+
{\Omega_{\rm N}^2\over 2}(x_1^2 +x_2^2) \biggr\} \biggr] \nonumber \\
& = &M\biggl[1+{1 \over c^2}\biggl({5W+U_{\rm i} \over M}+ {13M \over 4R}
+{1 \over 4R^3}(45\bI_{11}+8I_{\rm s})+O(R^{-5}) \biggr)\biggr]. 
\eeqn
The advantage of the PPN mass is that if we use it, $\Omega^2$ can be 
denoted only by $M_{\rm PPN}$ for the limit 
$I_{kk}/R^3 \rightarrow 0$\cite{will}. 
However, when $I_{kk}$ is large enough, it is not the 
conserved quantity any more, and not so useful. 

We define the center of mass by the conserved mass density as
\beq
x^i_{*} ={1 \over M_{*} }\int d^3x \rho_* x^i,
\eeq
and the $x_1$ coordinate of the center of mass for star 1 becomes
\beq
{1 \over c^2}\biggl(-{2I_{11} \over R^2}+O(R^{-4})\biggr).\label{ggeq}
\eeq
Note that in the PPN formalism\cite{will}, 
the center of mass of each star is defined as 
\beq
x^i_{\rm PPN } ={1 \over M_{\rm PPN} }
 \int d^3x \rho x^i \biggl[1+{1 \over c^2}
\biggl\{ {v^2 \over 2}+\varepsilon+3U-{U^{1\rightarrow 1} \over 2}+
{\Omega_{\rm N}^2\over 2}(x_1^2+x_2^2) \biggr\} \biggr], 
\eeq
and the results are the same. From Eq.(\ref{ggeq})
we define the orbital separation as 
\beq
R_{*}=R_{\rm PPN}=R\biggl\{1+{1 \over c^2}\biggl(
-{4I_{11} \over R^3}+O(R^{-5}) \biggr)\biggr\}.
\eeq

\section{The ellipsoidal approximation}

In this section, we introduce the ellipsoidal 
approximation for the density configuration 
proposed by Lai, Rasio and Shapiro(LRS)\cite{LRS}. 
In this approximation, the configuration of each star is assumed to be 
the ellipsoidal figure of its axial length $a_1$, $a_2$ and $a_3$. 
Then, the density profile is assumed to be 
\beq 
\rho=\rho_{\rm p}(\xi), \hskip 1cm {\rm for}~~
\xi=\sqrt{{x_1^2 \over \hat a_1^2}+{x_2^2 \over \hat a_2^2}
+{x_3^2 \over \hat a_3^2}}
\leq \xi_1~~{\rm and}~~\hat a_k={a_k \over \xi_1},
\eeq
where $\rho_{\rm p}(\xi)$ has the density profile of the spherical 
polytrope, and $\rho=0$ for $\xi > \xi_1$. 
Because of the polytropic relation, $P$ and $\varepsilon$ are 
also assumed to depend only on $\xi$. 

Here, there is no solid basis to justify this model: i.e., 
in this approximation, we do not rigidly treat (a) the change of 
the density profile due to the PN gravity 
which occurs even for the spherical star, 
and (b)the deformation of the star 
due to the tidal gravity by the companion star. 
These facts mean that several quantities of the system 
calculated below such as the energy, the angular momentum, the central 
density of the star, and so on, deviate systematically from their 
exact values(see section VI). 
However, (1) in the Newtonian case, this treatment is almost exact for the 
incompressible fluid\cite{Ch69}\cite{LRS}, and (2) 
in a previous paper\cite{TS}, we find that the tidal deformation 
by the PN gravity is a small 
effect for the incompressible fluid star. Thus, we may expect that the 
ellipsoidal approximation is good for a sufficiently stiff EOS(small $n$). 

Following the standard procedure, we set 
\beq
\rho_{\rm p}=\rho_{\rm c}\theta(\xi)^n,
\eeq
where $\rho_{\rm c}$ is the central density, and $\theta$ obeys the Lane-Emden 
equation\cite{shapiro}. Then, $M$, $I_{kk}$, $U_{\rm i}$, and 
$I_{\varepsilon kk}$ are immediately calculated as 
\beqn
&&M=4\pi\rho_{\rm c} a_1 a_2 a_3\xi_1^{-1}|\theta'|,\label{maseq}\\
&&I_{kk}={M a_k^2 \over 5}\kappa_n, \hskip 26mm \kappa_n \equiv 
{5 \over 3\xi_1^4 |\theta'|}\int_0^{\xi_1} \theta^n\xi^4 d\xi, \\
&&U_{\rm i}=k_1 K \rho_{\rm c}^{1/n}M, \hskip 22mm 
k_1\equiv {n(n+1) \over 5-n}\xi_1|\theta'|,\\
&&I_{\varepsilon kk}=5 n K\rho_{\rm c}^{1/n} I_{kk} {\sigma_n \over \kappa_n},
\hskip 1.2cm \sigma_n \equiv
{1 \over 3\xi_1^4|\theta'|}\int_0^{\xi_1} \theta^{n+1}\xi^4 d\xi , 
\eeqn
where $\theta'=\pa \theta/\pa \xi$ at $\xi_1$. 

As for $W$ and ${\cal M}_{kk}$, we follow LRS\cite{LRS} and set as 
\beqn
&&W={2\pi \bar \rho A_0 \over 5-n}M,\\
&&{\cal M}_{kk}=-{2\pi \bar \rho A_k a_k^2  \over 5-n}M,
\eeqn
where $\bar \rho=3M/(4\pi a_1a_2a_3)$, in order 
to guarantee that in the case of 
the incompressible ellipsoid or the spherical star, these 
are exact definitions. Due to the same principle, $I_{Ukk}$ is written as
\beq
I_{Ukk}={5-n \over 3}{W \over M} I_{kk}
\biggl( {5\sigma_n \over \xi_1|\theta'|\kappa_n}+1\biggr),
\eeq
where we make use of the fact that in the case of the spherical star 
and/or the incompressible star, $U^{1 \rightarrow 1}$ is written as
\beq
U^{1 \rightarrow 1}={5-n \over 3}{W \over M}\biggl(1+{\theta \over \xi_1 
|\theta'|}\biggr).
\eeq
Thus, all quantities to calculate $\Omega^2$ and $M_*$ 
reduce to the available form, and they are rewritten as
\beqn
\Omega^2&=&{M \over a_1^3}\biggl( {2 \over \hat R^3}+
{6\kappa_n q_{11} \over 5\hat R^5} \biggr)
+{M^2 \over c^2a_1^4}\biggl[{2 \over \hat R^3}\biggl\{5\hat W+\hat U_{\rm i}
-{9 \over 4\hat R}-{\kappa_n \over 10 \hat R^3}
(28-14\alpha_2^2-9\alpha_3^2)\biggr\}\nonumber\\
&&\hskip 5cm ~+{6\kappa_n q_{11} \over 5\hat R^5}\hat W
\biggl({23-n \over 6}+\gamma_n{15n+35 \over 6(n+1)} \biggr)\biggr]\nonumber\\
&\equiv & {M \over a_1^3}\hat \Omega_{\rm N}^2(\hat R)
+{M^2\over c^2a_1^4}\hat \Omega_{\rm PN}^2(\hat R),
\label{omefinal} \\
M_*&=&M\biggl[1+{ M \over c^2 a_1}\biggl\{6\hat W+{13 \over 4\hat R}
+{\kappa_n \over 20\hat R^3}(15q_{11}+4q_{\rm s})\biggr\}\biggr],
\label{masfinal}
\eeqn
where $q_{11}=2-\alpha_2^2-\alpha_3^2$, $q_{\rm s}=1+\alpha_2^2$, 
$\hat R=R/a_1$, $\gamma_n=(5-n)\sigma_n/(\xi_1|\theta'|\kappa_n)$, and
\beqn
\hat W(\hat R)&=&
W{a_1 \over M^2}={3\hat A_0 \over 2(5-n)\alpha_2\alpha_3},\\
\hat U_{\rm i}(\hat R)&=&U_{\rm i}{a_1 \over M^2}=
{n \over 3}\biggl(\hat W -{\kappa_n q_{11} \over 5\hat R^3}
-{\kappa_n q_{\rm s} \over 5}\hat \Omega_{\rm N}^2\biggr).
\eeqn
Here, to derive the expression for $\hat U_{\rm i}$, 
we use the scalar virial relation in the Newtonian order. 
In the ellipsoidal approximation, $\alpha_2$ and 
$\alpha_3$, which are functions of $\hat R$, 
are assumed to be determined from the 
tensor virial relations in the Newtonian order as\cite{LRS}
\beqn
&&{2\pi \bar \rho A_1 a_1^2  \over 5-n}M={1 \over n} k_1 K \rho_{\rm c}^{1/n}M
+{2MI_{11} \over R^3}+ \Omega_{\rm N}^2 I_{11},\nonumber \\
&&{2\pi \bar \rho A_2 a_2^2  \over 5-n}M={1 \over n}k_1K\rho_{\rm c}^{1/n}M
-{MI_{22} \over R^3}+\Omega_{\rm N}^2I_{22},\nonumber \\
&&{2\pi \bar \rho A_3 a_3^2  \over 5-n}M={1 \over n}k_1K\rho_{\rm c}^{1/n}M
-{MI_{33} \over R^3}. \label{tenvil}
\eeqn 

\section{The energy and the angular momentum}

Next, we consider the total energy of the system. 
The energy for each star is calculated from $E_1=E_{\rm N}
+E_{\rm PN}/c^2$, where\cite{chand}\cite{shibapn}
\beqn
E_{\rm N}&=&
\int \rho \biggl(\varepsilon+{1 \over 2}v^2-{1 \over 2}U\biggr)d^3x,\\
E_{\rm PN}&=&\int \rho \biggl(
{5 \over 8}v^4+{5 \over 2}v^2U+\Gamma\varepsilon v^2
+2\varepsilon U
-{5 \over 2}U^2+{1 \over 2}\beta_{\varphi}\Omega^2 \biggr) d^3x,
\eeqn
and $v^2=\varpi^2 \Omega^2$. 
Even for the compressible PN Darwin problem, the expression of 
$E_{\rm N}$ is soon derived by using the quantities defined above as 
\beqn
E_{\rm N}&=&U_{\rm i}-W+{\Omega^2 \over 2}\biggl({MR^2 \over 4}+
I_{\rm s} \biggr)-{M \over 2}\biggl({M \over R}+{3\bI_{11} \over R^3}\biggr)
\nonumber \\
&=&\biggl({M^2 \over a_1}\biggr)\biggl[\hat U_{\rm i}-\hat W
+{\hat \Omega^2 \over 2}\biggl({\hat R^2 \over 4}
+{\kappa_n q_{\rm s} \over 5} \biggr)
-{1 \over 2}\biggl({1 \over \hat R}+{\kappa_n q_{11} \over 5 \hat R^3}
\biggr)\biggr] \nonumber \\
&\equiv &\biggl({M^2 \over a_1}\biggr)\biggl[
\hat E_{\rm N}(\hat R)
+{M \over 2a_1c^2}\hat \Omega_{\rm PN}^2\biggl({\hat R^2 \over 4}
+{\kappa_n q_{\rm s} \over 5} \biggr)\biggr],\label{eneneq}
\eeqn
where $\hat \Omega^2=\Omega^2 a_1^3/M
=\hat \Omega_{\rm N}^2+(M/c^2a_1)\hat \Omega_{\rm PN}^2$. 
On the other hand, to express $E_{\rm PN}$, we are required to 
evaluate new integrals as 
\beqn
J_1&=&\int \rho\varepsilon U^{1 \rightarrow 1} d^3x,\\
J_2&=&\int \rho (U^{1 \rightarrow 1})^2 d^3x,\\
J_3&=&\int \rho \varpi^2 U^{1 \rightarrow 1} d^3x,\\
J_4&=&\int \rho U^{1 \rightarrow 1}U^{2 \rightarrow 1} d^3x,\\
J_5&=&\int \rho \beta_{\varphi}^{1 \rightarrow 1} d^3x. 
\eeqn
To evaluate $J_1 \sim J_4$, we use the same principle in 
calculating $\bI_{U11}$. Then, they become
\beqn
J_1&=& {n \over 3}K\rho_{\rm c}^{1/n}W \xi_1|\theta'|
\biggl({(5-n)\lambda_n \over \xi_1^2|\theta'|^2}+n+1 \biggr),\\
J_2&=&{5-n \over 9}{W^2 \over M}\biggl({(5-n)\lambda_n \over \xi_1^2
|\theta'|^2}+n+7\biggr) ,\\
J_3&=&{R^2 W \over 2}+{5-n \over 3}
{W \over M}I_{\rm s}
\biggl(5{\sigma_n \over \xi_1|\theta'|\kappa_n}+1 \biggr),\\
J_4&=&2W\biggl({M \over R}+{3\bI_{11} \over 2R^3}\biggr)
+{\bI_{11} \over 2R^3}(5-n)W 
\biggl( 5{\sigma_n \over \xi_1|\theta'|\kappa_n}+1 \biggr)+O(R^{-5}),
\eeqn
where
\beq
\lambda_n={1 \over \xi_1^2|\theta'|}\int_0^{\xi_1} \theta^{n+2}\xi^2d\xi.
\eeq
On the other hand, $J_5$ is written as
\beqn
J_5&=&{R^2 \over 4}({\cal M}_{22}-7W)
-{7 \over 2}\int \rho (D_1x_1+D_2x_2) d^3x \nonumber \\
&&-{1 \over 2}\int \rho
(x_2^2D_{1,1}+x_1^2D_{2,2}-x_1x_2D_{1,2}-x_1x_2D_{2,1}) d^3x.
\eeqn
Unfortunately, the integrand in 
the second and third terms cannot be expressed by $\theta$. 
Hence, we need to introduce a new function: 
To evaluate $D_1$ and $D_2$, we assume that the density profile $\rho$, 
which appears in the source terms for their Poisson equations, is equal to 
that of the spherical polytrope $\rho_{\rm p}(\xi)$. Then, we can
write $D_k$ as 
\beq
D_k=4\pi\rho_{\rm c} \hat a^2\phi(\xi) x_k,
\eeq
where $\hat a^3=\hat a_1 \hat a_2 \hat a_3$ and $\phi$ satisfies
\beq
{d^2 \phi \over d\xi^2}+{4 \over \xi}{d \phi \over d\xi}=-\theta^n.
\eeq
Note that the following relation holds;
\beq
\sum_k D_{k,k}=U^{1\rightarrow 1}+\sum_k x_k U^{1\rightarrow 1}_{,k}. 
\eeq
Also, for the spherical star, 
\beq
4\pi\rho_{\rm c} \hat a^2={5-n \over 3\xi_1|\theta'|}{W \over M}.
\eeq
>From these two equations, we have the relation
\beq
3\phi+\xi{d\phi \over d\xi}
=\xi_1|\theta'|+\theta+\xi{d\theta \over d\xi}.\label{phixi}
\eeq
Thus, $\phi$ is solved under the boundary conditions 
$\phi=(\xi_1|\theta'|+1)/3$ and $d\phi/d\xi=0$ at $\xi=0$. 
Using this definition of $D_k$, $J_5$ can be written as 
\beq
J_5={R^2 \over 4}({\cal M}_{22}-7W)
-{4M^2 \over 3(a_1a_2a_3)^{1/3}}(a_1^2+a_2^2)\eta_n,
\eeq
where
\beq
\eta_n={1 \over \xi_1^5 |\theta'|^2} \int_0^{\xi_1} 
\theta^n \phi \xi^4 d\xi .  
\eeq
Exchanging $M^2/(a_1a_2a_3)^{1/3}$ into $(5-n)W/3$, we finally get 
\beq
J_5={R^2 \over 4}({\cal M}_{22}-7W)
-{4W \over 9}(5-n)(a_1^2+a_2^2)\eta_n. 
\eeq

Gathering the above results, we obtain the expression for $E_{\rm PN}$ as 
\beqn
E_{\rm PN}&=& -{5(5-n) \over 18}{W^2 \over M}
\biggl(\tilde \lambda_n+n+7\biggr)
+{2(5-n) \over 3(n+1)}{U_{\rm i}W \over M}
\biggl(\tilde \lambda_n+n+1\biggr)\nonumber \\
&&+{M\over R}\biggl\{-{37 \over 4}W+\biggl({5n \over 2}+{1 \over 2}\biggr)
{U_{\rm i} \over n}+{{\cal M}_{22} \over 4}\biggr\}-{7M^3 \over 32R^2} 
\nonumber \\
&&+{1 \over R^3} \biggl[ W \biggl\{ -\bI_{11}\biggl( {83-10n \over 4}
+{25 \over 2}\gamma_n \biggr)+I_{\rm s} \biggl( {25 \over 3} \gamma_n +
{5(5-n) \over 3}-{20(5-n) \over 9}{\eta_n \over \kappa_n}\biggr)\biggr\}
\nonumber \\
&&\hskip 1cm +{9 \over 4}\bI_{11}{\cal M}_{22}+
{U_{\rm i} \over n}\biggl\{ \biggl( {15n \over 2}+{9 \over 2} \biggr)\bI_{11}
+{5\gamma_n  \over n+1}\Bigl( 2(n+1) I_{\rm s}+3n\bI_{11}\Bigr) \biggr\}\biggr]
\nonumber \\
&&+{M^2 \over 16R^4}\Bigl(106I_{11}+25I_{22}-71I_{33}\Bigr),
\eeqn
where $\tilde \lambda_n=(5-n)\lambda_n/(\xi_1^2|\theta'|^2)$. 
Using the scalar and tensor virial relations in the Newtonian order, 
$E_{\rm PN}$ is rewritten to a simple form 
\beqn
E_{\rm PN}&=& -{5(5-n) \over 18}{W^2 \over M}
\biggl(\tilde \lambda_n+n+7\biggr)
+{2(5-n) \over 3(n+1)}{U_{\rm i}W \over M}
\biggl(\tilde \lambda_n+n+1\biggr)\nonumber \\
&&+{M\over R}\biggl( -{55 \over 6}W+{5 \over 2}U_{\rm i}
\biggr)-{7M^3 \over 32R^2} \nonumber \\
&&+{W \over R^3} \biggl[ \bI_{11}\biggl(5n -20+{5n \over n+1}\gamma_n
-{25 \over 2}\gamma_n \biggr)+I_{\rm s} \biggl( {35 \over 3} \gamma_n +
{5(5-n) \over 3}-{20(5-n) \over 9}{\eta_n \over \kappa_n}\biggr)\biggr]
\nonumber \\
&&+{M^2 \over 48R^4}\Bigl(302I_{11}+59I_{22}-209I_{33}\Bigr)+O(R^{-5})
\nonumber \\
&\equiv &\biggl({M^3 \over a_1^2}\biggr)
\hat E_{\rm PN}(\hat R), \label{enepneq}
\eeqn
where 
\beqn
\hat E_{\rm PN}&=&
 -{5(5-n) \over 18}\hat W^2\biggl(\tilde \lambda_n+n+7\biggr)
+{2(5-n) \over 3(n+1)}\hat U_{\rm i} \hat W
\biggl(\tilde \lambda_n+n+1\biggr)\nonumber \\
&&+{1\over \hat R}\biggl( -{55 \over 6}\hat W+{5 \over 2}\hat U_{\rm i}
\biggr)-{7 \over 32\hat R^2} \nonumber \\
&&+{\kappa_n\hat W \over 15\hat R^3} 
\biggl\{ q_{11}\biggl(5n -20+{5n \over n+1}\gamma_n
-{25 \over 2}\gamma_n \biggr)+q_{\rm s} \biggl( 35 \gamma_n +
5(5-n)-{20(5-n) \over 3}{\eta_n \over \kappa_n}\biggr)\biggr\}
\nonumber \\
&&+{\kappa_n \over 240\hat R^4}\Bigl(302+59\alpha_2^2-209\alpha_3^2\Bigr)
+O(R^{-5}). 
\eeqn
For the incompressible case $n=0$, $\kappa_0=1$, $\sigma_0=2/35$, 
$\gamma_0=1/7$, $\eta_0=6/35$, $\lambda_0=8/35$(see Table I), 
$U_{\rm i}=0$, and 
$W/M=2\pi\rho A_0/5$, the expression of $E_{\rm PN}$, of course, 
agrees with that obtained in a previous paper\cite{TS}. 

The angular momentum for each star 
$J_1 = J_{{\rm N}} +J_{{\rm PN}}/c^2$ is calculated from 
\beqn
J_{\rm N} &=& \int \rho v_{\varphi} d^3 x, \nonumber \\
J_{\rm PN} &=& \int \rho \left[ v_{\varphi} 
\left( \Omega_{\rm N}^2 \varpi^2 +6U +\Gamma\varepsilon \right) 
+ \beta_{\varphi} \Omega \right] d^3 x,
\label{angeq}
\eeqn
where $v_{\varphi} = \Omega \varpi^2$. 
Using $J_1 \sim J_5$, they are soon expressed as 
\beqn
J_{\rm N} &=& \Omega \left( {MR^2 \over 4} + I_{\rm s} \right)
=M^{3/2}a_1^{1/2}\hat \Omega\biggl({\hat R^2 \over 4}
+{\kappa_n q_{\rm s} \over 5}\biggr)
\equiv M^{3/2}a_1^{1/2} \hat \Omega 
\hat J_{\rm N}(\hat R), 
\label{angneq}\\
J_{\rm PN} &=&\Omega_{\rm N} \biggl[{R^2 \over 4}(U_{\rm i} + 5W)
+{5 \over 2}RM^2+I_{\rm s}{W \over M}\biggl\{ {35\gamma_n \over 3}
+2(5-n)-{20 \over 9}(5-n){\eta_n \over \kappa_n} \biggr\}\nonumber\\
&& \hskip 2cm +{M \over 4R}(20I_{11}+15I_{22}-11I_{33})\biggr] \nonumber\\
&=&M^{5/2}a_1^{-1/2}\hat \Omega_{\rm N}
\biggl[{\hat R^2 \over 4}(\hat U_{\rm i} + 5\hat W)
+{5 \over 2}\hat R+{\kappa_n q_{\rm s}\hat W \over 5}\biggl\{
{35\gamma_n \over 3}
+2(5-n)-{20 \over 9}(5-n){\eta_n \over \kappa_n} \biggr\}\nonumber\\
&& \hskip 3cm +{\kappa_n \over 20\hat R}(20+15\alpha_2^2-11\alpha_3^2)\biggr]
\nonumber \\
&\equiv &M^{5/2}a_1^{-1/2} \hat \Omega_{\rm N} 
\hat J_{\rm PN}(\hat R). 
\label{angpneq}
\eeqn

Once we determine $\alpha_2(\hat R)$ and $\alpha_3(\hat R)$ from 
Eqs.(\ref{tenvil}), we can calculate 
$\hat \Omega_{\rm N}$, $\hat \Omega_{\rm PN}$ 
$\hat E_{\rm N}$, $\hat E_{\rm PN}$, $\hat J_{\rm N}$, and $\hat J_{\rm PN}$ 
as a function of $\hat R$ 
right now using the numerical values for $\xi_1$, $|\theta'|$, 
$k_1$, $\kappa_n$, $\sigma_n$, $\lambda_n$ and $\eta_n$ shown in 
Table I. 
To evaluate the angular velocity, the energy and the 
angular momentum, however, we further need the equation for determination 
of $a_1$. We will derive it in the next section. 

\section{Scalar virial relation in the post-Newtonian approximation}

Up to this section, we have derived the equations required 
for determining the equilibrium sequence 
except for $a_1$(or $\rho_{\rm c}$). Here, we derive 
the equation for $a_1$(or $\rho_{\rm c}$) in the PN order. 
Such an equation should come from the scalar virial relation in the 
PN approximation, which is derived by 
\beqn
& &\sum_i \int d^3x \rho x_i{\pa \over \pa x_i}\biggl[
K'\rho^{\Gamma-1}-{1 \over 2c^2}\biggl(K'\rho^{\Gamma-1}\biggr)^2\biggr]
\nonumber \\
&=&\sum_i \int d^3x \rho x_i{\pa \over \pa x_i}\biggl[U-{X_0 \over c^2}
+\biggl\{{\varpi^2 \over 2}+{1 \over c^2}\Bigl(2\varpi^2U-X_{\Omega}
+ \beta_{\varphi}\Bigr)\biggr\}\Omega^2
+{\varpi^4 \over 4c^2}\Omega^4\biggr]. \label{pntv1}
\eeqn
Using the equations shown in previous sections, 
the left-hand side of Eq.(\ref{pntv1}) can be written as
\beq
M\biggl[-{3 \over n}k_1K\rho_{\rm c}^{1/n}+{1 \over c^2}
{3(n+1)^2 \over n+2}\lambda_nK^2
\rho_{\rm c}^{2/n}\biggr].
\eeq
In the right-hand side of Eq.(\ref{pntv1}), 
many terms are soon integrated as  
\beqn
-W+{3M \over R^3}\bI_{11}&&+\Omega^2 I_{\rm s} +{1 \over c^2}\biggl[
-K_1 +\biggl({15+n \over R^3}W
+{6M^2 \over R^4}\biggr) \bI_{11}\nonumber \\
&&+\Omega_{\rm N}^2\biggl\{ 2\sum_i ({\cal M}_{ii11}+{\cal M}_{ii22})
-{R^2 \over 2}W
+4(I_{U11}+I_{U22})-K_2+K_3  \nonumber \\
&&\hskip 1cm +{M \over 8R}(2I_{11}+15I_{22}-15I_{33})\biggr\}
+{\Omega_{\rm N}^4 \over 4}R^2(3I_{11}+I_{22}) +O(R^{-5})\biggr], 
\eeqn
but for 
\beqn
K_1&=&\sum_i\int \rho X_{0,i}^{1\rightarrow 1}x_i d^3x,\\
K_2&=&\sum_i\int \rho X_{\Omega,i}^{1\rightarrow 1}x_i d^3x,\\
K_3&=&\sum_i\int \rho \beta_{\varphi,i}^{1\rightarrow 1} x_i d^3x. 
\eeqn
$K_1$ is evaluated as 
\beqn
K_1&=&-\int\int d^3x d^3y\rho(x)\rho_X^{1\rightarrow 1}(y) \sum_i x_i
{\pa \over \pa x_i}\biggl({1 \over |{\bf x}-{\bf y}|} \biggr)\nonumber \\
&=&\int\int d^3x d^3y\rho(x)\rho_X^{1\rightarrow 1}(y) \sum_i x_i
{\pa \over \pa y_i}\biggl({1 \over |{\bf x}-{\bf y}|} \biggr)\nonumber \\
&=&\int d^3y\rho_X^{1\rightarrow 1}(y) \sum_i D_{i,i} \nonumber \\
&=&\int d^3y\rho_X^{1\rightarrow 1}(y) 
(U^{1\rightarrow 1}+\sum_i y_iU^{1\rightarrow 1}_{,i})\nonumber\\
&=&\biggl(1+{3\over n}\biggr)(J_1+J_1')+2(J_2+J_2'+J_4+J_4'),
\eeqn
where we define 
\beqn
J_1'&=&\sum_i\int \rho U_{,i}^{1\rightarrow 1}x_i \varepsilon d^3x,\\
J_2'&=&\sum_i\int \rho U_{,i}^{1\rightarrow 1}x_i 
U^{1\rightarrow 1} d^3x,\\
J_4'&=&\sum_i\int \rho U_{,i}^{1\rightarrow 1}x_i 
U^{2\rightarrow 1} d^3x. 
\eeqn
In the ellipsoidal approximation, they are written as 
\beqn
J_1'&=&-{\tilde \lambda_n \xi_1|\theta'| \over n+2}WKn\rho_{\rm c}^{1/n},
\nonumber\\
J_2'&=&-{5-n \over 3(n+2)}{W^2 \over M}(\tilde \lambda_n+n+2),\nonumber \\
J_4'&=&-W\biggl({M \over R}+{3\bI_{11} \over 2R^3}
+{25 \gamma_n \over 2(n+1)}{\bI_{11} \over R^3}\biggr)+O(R^{-5}). 
\eeqn
Using the same method as that adopted in calculating $K_1$, 
\beqn
K_2&=&2\int d^3x \rho\biggl[\biggl(x_1^2+x_2^2+{R^2 \over 4}\biggr)
\Bigl(U^{1\rightarrow 1}+\sum_i x_iU^{1\rightarrow 1}_{,i}\Bigr)
\biggr] \nonumber \\
&=&{R^2 \over 2}W+2(I_{U11}+I_{U22})+2\sum_i ({\cal M}_{ii11}
+{\cal M}_{ii22}). 
\eeqn
$K_3$ is rewritten as
\beqn
K_3&=&-{1 \over 8}  \int  d^3x \rho \sum_i 
\biggl[R^2 x_i{\pa \over \pa x_i}\biggl(7U^{1\rightarrow 1}
+D_{2,2}-U^{1\rightarrow 1}_{,2}x_2\biggr) \nonumber \\
&& +4 x_i{\pa \over \pa x_i}\biggl(7x_1D_1+7x_2D_2 
+x_1^2D_{2,2}+x_2^2D_{1,1}-x_1x_2D_{1,2}-x_1x_2D_{2,1}\biggr)\biggr].
\eeqn
Using the same method as that in calculating $J_5$ with Eq.(\ref{phixi}), 
\beq
K_3={R^2 \over 8}(7W-{\cal M}_{22})
-{20 \over 9}{W \over M}I_{\rm s}\biggl[ {3(5-n) \over 5}+
{3(n-4) \over n+1} \gamma_n - {\eta_n \over \kappa_n}(5-n) \biggr]. 
\eeq

Gathering these results, the PN scalar virial relation is written as
\beq
{3k_1 \over n}K\rho_{\rm c}^{1/n}+{1 \over c^2}BK^2 \rho_{\rm c}^{2/n}
=C{M \over a_1}
+D{M^2 \over c^2a_1^2},\label{scalartv}
\eeq
where
\beqn
B&=&-{3(n+1)^2 \over n+2}\lambda_n ,\nonumber\\
C&=&\hat W-{\kappa_n q_{11}\over 5\hat R^3} - {\hat \Omega_{\rm N}^2
\kappa_n q_{\rm s} \over 5}={3 \over n}\hat U_{\rm i},\nonumber\\
D&=&{(n+3)(5-n) \over 3(n+1)n}\hat W \hat U_{\rm i}\biggl({n-1 \over n+2}
\tilde \lambda_n+n+1  \biggr)
+{2(5-n) \over 9}\hat W^2\biggl({n-1 \over n+2}\tilde \lambda_n+n+4\biggr)
\nonumber\\
&&+{13 \over 6\hat R}\hat W+{\kappa_n \hat W\over 15\hat R^3}
\biggl[q_{11}\biggl(-{11 \over 2}
-2n+{5(n-4) \over n+1}\gamma_n\biggr)\nonumber \\
&& \hskip 4cm -{4 \over 3}q_{\rm s}\biggl({135-15n \over n+1}\gamma_n
+{9n +15\over 2}+{10\eta_n \over \kappa_n}(5-n) \biggr)\biggr]\nonumber \\
&& -{\kappa_n \over 60\hat R^4}(86+35\alpha_2^2-68\alpha_3^2). 
\eeqn
Using Eqs.(\ref{maseq}) and (\ref{scalartv}), 
$a_1=a_{1{\rm N}}+\delta a_1/c^2$ is determined as 
\beqn
a_{1{\rm N}}&=& \biggl( {\hat U_{\rm i} \over k_1K} \biggr)^{n/(n-3)}
\biggl( {\xi_1 \over 4\pi\alpha_2\alpha_3|\theta'|} \biggr)^{1/(3-n)}
M_{\rm N}^{(n-1)/(n-3)},\nonumber \\
\delta a_1&=&a_{1{\rm N}}\biggl[{1-n \over 3-n}{\delta M \over M_{\rm N}}
+{n \over C(n-3)}{M_{\rm N} \over a_{1{\rm N}} }\biggl\{
D-B\biggl( {\hat U_{\rm i} \over k_1}  \biggr)^{2}\biggr\}
\biggr],\label{aaeq}
\eeqn
where we split $M$ as $M_{\rm N}+\delta M/c^2$. 
The first term of the equation of $\delta a_1$ originates from 
the Newtonian property, and the second and third terms(which include 
$B$ and $D$) are purely the PN origin. 
Since $B < 0$ and $D > 0$ for $0< n <3$, 
the PN effect tends to decrease $\delta a_1 $ 
for realistic neutron stars, and as a result makes 
the coordinate radius of the star smaller than that in the Newtonian case. 

\section{Analysis of equilibrium sequences} 

An equilibrium sequence of the compressible Darwin ellipsoids is 
determined fixing $M_*$ and $K$. The procedure is as follows.

\noindent
(1) using Eqs.(\ref{tenvil}), we calculate the equilibrium sequence 
in the Newtonian order. Up to this stage, $\alpha_2(\hat R)$ and 
$\alpha_3(\hat R)$ are determined, and the results are checked by 
comparing those by LRS\cite{LRS}. 

\noindent
(2) $M$ is determined by the condition $M_*=$constant. From 
Eq.(\ref{masfinal}), 
\beq
M=M_*\biggl[1-{M_* \over c^2a_{1{\rm N}}}\biggl\{6\hat W+{13 \over 4\hat  R}
+{\kappa_n \over 20 \hat  R^3}(15q_{11}+4q_{\rm s})\biggr\}
\biggl].\label{mmeq}
\eeq
Then, we set $M_{\rm N}=M_*=$constant and 
\beq
\delta M=-M_*
{M_* \over a_{1{\rm N}} }\biggl\{6\hat W+{13 \over 4\hat  R}
+{\kappa_n \over 20 \hat  R^3}(15q_{11}+4q_{\rm s})\biggr\}.
\eeq

\noindent
(3) after substituting Eqs.(\ref{aaeq}) and (\ref{mmeq}) into the 
equations for $\Omega^2$, $E_1$ and $J_1$, we rewrite them as 
\beqn
\Omega^2&=&{M_* \over a_{1\rm N}^3}\biggl[\hat \Omega_{\rm N}^2
+{1 \over c^2}\biggl\{ \biggl({\delta M \over M_*}-
{3\delta a_1 \over a_{1\rm N} }\biggr)\hat \Omega_{\rm N}^2
+{M_* \over a_{1\rm N}}\hat \Omega_{\rm PN}^2\biggr\}\biggr],\\
E_1&=&{M_*^2 \over a_{1N}}\biggl[\hat E_{\rm N}+{1 \over c^2}
\biggl\{\biggl({2\delta M \over M_*}-{\delta a_1 \over a_{1{\rm N}} }\biggr)
\hat E_{\rm N}+{M_* \over 2a_{1{\rm N}}}\hat \Omega_{\rm PN}^2
\biggl({\hat R^2 \over 4}+{\kappa_n q_{\rm s} \over 5}\biggr) 
+{M_* \over a_{1{\rm N}} }\hat E_{\rm PN}\biggr\}\biggr],\\
J_1&=&M_*^{3/2}a_{1{\rm N}}^{1/2}\hat \Omega_{\rm N}
\biggl[ \hat J_{\rm N}+{1 \over c^2}\biggl\{ \biggl( {3\delta M \over 2M_*}
+{\delta a_1 \over 2a_{1\rm N} } 
+{M_* \over a_{1\rm N}}{\hat \Omega_{\rm PN}^2 \over 2\hat \Omega_{\rm N}^2}
\biggr)\hat J_{\rm N}
+{M_* \over a_{1\rm N} }\hat J_{\rm PN} \biggr\}\biggr],
\eeqn
and define the total energy and the total angular momentum of the system as 
$E=2E_1$ and $J=2J_1$. 
Then, using  $\alpha_2(\hat R)$ and $\alpha_3(\hat R)$ obtained at (1), 
we calculate the PN sequences of $\Omega$, $E$, and $J$ as a 
function of $\hat R$. 

\noindent
We repeat these procedures changing the compactness of each star 
of binary defined as $C_{\rm s} = M_*/c^2a_*$, where $a_*$ is the 
radius of the Newtonian spherical star 
\beq
a_*=M_*^{(n-1)/(n-3)}\xi_1\biggl({n+1 \over 4\pi}K\biggr)^{n/(3-n)}
(4\pi\xi_1^2|\theta'|)^{-(1-n)/(3-n)}.
\eeq
For the following, we also define 
\beq
\rho_*={3M_* \over 4\pi a_*^3}.
\eeq

In figs.1 and 2, we show the energy($E/(M_*^2/a_*)$), 
the angular momentum($J/\sqrt{M_*^3a_*}$), and the angular 
velocity($\Omega/\sqrt{\pi\rho_*}$) as a function of the orbital separation, 
$R_*/a_*$ for $n=0.5$(figs.1) and 1(figs.2), respectively. 
In figs.1, 
the solid and dotted lines denote the results calculated by 
the ellipsoidal approximation for $C_{\rm s}=0$ and 0.02, 
and the open and filled circles denote the results by the numerical 
calculation\cite{shibapn} for $C_{\rm s}=0$ and 0.02, respectively. 
In figs.2, the solid and dotted lines denote the results calculated by
the ellipsoidal approximation for $C_{\rm s}=0$ and 1/60, 
and the open and filled circles denote the results by the numerical
calculation for $C_{\rm s}=0$ and 1/60, respectively. 
These figures indicate that 
the ellipsoidal approximation works fairly well for calculating 
the angular momentum and the angular velocity, but does not very well 
for calculating the energy: The energy calculated 
by the ellipsoidal approximation systematically deviates from the 
numerical results. This is because 
we do not rigidly take into account the change of the 
density profile of the star 
due to the PN gravity even for $R \rightarrow \infty$, although the energy 
is sensitive to the density configuration(i.e., $\hat W$ or $\hat A_0$). 
The angular momentum and the angular velocity are not sensitive to the 
density profile, but to the global quantities 
such as the mass and the spin as it can be found from their equations. 
However, the deviation of the energy is systematic, so that this weak point 
could be improved if we artificially reduce 
$\hat A_0$ by a small correction 
factor which will depend on $C_{\rm s}$ and $n$\footnote{
For the same $C_{\rm s}$, 
we will need a larger correction factor for larger $n$, and for $n=0$, 
we do not need the correction.}. 

The ellipsoidal approximation is not quantitatively a 
sophisticated model for calculating the various quantities, in particular 
the energy, for close binaries of $\hat R \sim 3$. 
Nevertheless, it is a qualitatively good model: 
Figs.1 and 2 indicate that the coordinate separation at the ISCCO 
becomes small due to the PN effect\footnote{
Since we use the ellipsoidal approximation for the density profile 
which is not the strict one, 
and also neglect sufficiently high order terms of $R^{-1}$ 
in the energy and the angular momentum, the location of the 
energy minimum does not coincide with that of the angular momentum minimum. 
However, the difference is not large like the results in a 
previous paper\cite{TS}.}, and this feature agrees 
with the previous numerical results\cite{shibapn}. 
We show the relation between the angular momentum and the 
angular velocity for $n=0.5$ and 1.0 in figs.3. These indicate that 
$\Omega$ at the ISCCO increases with the PN effect, and this feature 
also agrees with the numerical results\cite{shibapn}. 

In figs.4$-$8, we show the angular momentum as a function of $R_*$(a) 
and the angular velocity as a function of the angular momentum(b) for 
$n=0.25$(figs.4), 0.5(figs.5), 0.75(figs.6), 1.0(figs.7), 1.5(figs.8) and 
for $C_{\rm s}=0$(solid line), 0.01(dotted line), 
0.02(dashed line), 0.03(long dashed line), respectively.  
We can see the feature found in figs.1$-$3 again; i.e., (1) the coordinate 
radius at the ISCCO decreases with the increase of $C_{\rm s}$, and (2) 
the orbital angular velocity at the ISCCO increases with $C_{\rm s}$. 
Thus, we conclude that such features hold irrespective of the EOS 
if the ISCCO exists at the Newtonian order. 
These figures also indicate that the star of a softer EOS is more 
susceptible to the PN effect, and as a result the 
orbital radius and the angular velocity at the ISCCO change much. 

Finally, we argue the stability of each star against the 
collapse due to the influence 
of the GR gravity from the companion star considering the central density 
$\rho_{\rm c}$. From Eqs.(\ref{maseq}), (\ref{aaeq}), and (\ref{mmeq}), we 
obtain 
\beqn
\rho_{\rm c}&=&\rho_{\rm N}\biggl[ 1 + {1 \over c^2}\biggl(
{\delta M \over M_*}-{3\delta a_1 \over a_{1\rm N}} \biggr)\biggr]\nonumber\\
&=&\rho_{\rm N}\biggl[ 1+{M_* \over c^2a_{1\rm N}}\biggl\{
{n \over 3(3-n)}\biggl(19+4n-3n^2+{(3n-1)(5-n)\tilde \lambda_n \over n+1}
\biggr)\hat W \nonumber \\
&& \hskip 2cm
+ {A_3 \over \hat R^3}+{A_4 \over \hat R^4}+O(\hat R^{-5})\biggr\} \biggr], 
\label{rhoeq}
\eeqn
where 
\beq
\rho_{\rm N}={M_*\xi_1 \over 4\pi |\theta'|\alpha_2\alpha_3a_{1\rm N}^3},
\eeq
and $A_3$ and $A_4$ are functions of $\hat R$. 
Note that in $\hat W$, the effect of $O(\hat R^{-3})$ is implicitly 
included because $\alpha_2$ and $\alpha_3$ behave as $1+O(\hat R^{-3})$. 
The lowest order terms of $\hat R$(the terms of $O(\hat R^{0})$) are 
concerned in the well known GR instability to the radial collapse 
for each star itself\cite{chandra}: Since the coefficient 
of $\hat W$ is always positive for $0< n < 3$, in the case 
of $n \rightarrow 3$ or of sufficiently large $\hat W $, 
$\rho_{\rm c} a_*^2/c^2 \sim M_*/(c^2a_*)$ 
becomes larger than unity, which means 
that the star is swallowed inside its Schwarzschild radius to be 
a black hole due to the GR gravity. 

In Eq.(\ref{rhoeq}), the terms of $O(R^{-1})$ do not appear: 
As Wiseman points out, this is the consequence of the strong 
equivalence principle\cite{wiseman}. Thus, 
the effect due to the GR gravity of the companion star appears from the 
terms of $O(R^{-3})$ which are concerned in the tidal effect in the PN order. 
As Lai pointed out\cite{lai}, the Newtonian tidal effect 
tends to stabilize each star; i.e., it reduces the central density. 
Actually, $\pa \rho_{\rm N}/\pa R$ 
is always positive in the present case. 
Here, we demonstrate that the PN tidal 
effect also reduces the central density. 
In figs.9, we show $\rho_{\rm c}/\rho_{\rm N}$ as a 
function of the orbital separation for $n=0.5$(a), 1(b), and 1.5(c). 
In each figure, the solid, dotted, dashed, and long dashed lines 
denote the cases for $C_{\rm s}=0$, 0.01, 0.02 and 0.03, respectively. 
These figures indicate that the GR effect 
due to the companion star reduces the central density. This seems to 
mean that it will act to stabilize each star of binary. 
Recently, Wilson et al. calculated the equilibrium state of binary neutron 
stars using the semi-relativistic approximation\cite{wilson}, 
and they showed that the 
central density of each star increases with the decrease of the orbital 
separation for a critical orbital separation. However, 
the present analysis suggests that for the 
corotating binary neutron stars, such an effect cannot be 
explained by the first PN correction of general relativity.

\section{summary}

Using the ellipsoidal model for the
density configuration(the ellipsoidal approximation), 
we have calculated the equilibrium states of 
corotating binary stars of the stiff polytropic EOS 
in the first PN approximation of general relativity. 
Comparing the equilibrium sequences with those obtained  
by numerical calculation\cite{shibapn}, we confirm that 
the ellipsoidal approximation is 
a qualitatively good model to see the several PN effects 
in binary neutroOAn stars, and then get the conclusions as follows. 

\noindent
(1) Irrespective of the polytropic index $n$, the PN effects make the 
orbital separation at the ISCCO smaller 
and the orbital angular velocity at the ISCCO larger. 

\noindent
(2) As in the Newtonian case\cite{lai}, the PN tidal effect from the 
companion star reduces the central density of each star. Thus, it seems 
to act the stabilization of each star. 

These results will be helpful to check numerical results in 
PN and/or full GR numerical computations 
of obtaining the equilibrium states of binary neutron stars. 

\acknowledgments

Numerical computations were partly 
performed on FACOM VPP-300 in data processing center of National Astronomical
Observatory in Japan. This work was in part supported by 
Japanese Grant-in-Aid of Ministry of Education, Culture, 
Science and Sports (Nos. 08237210 and 08NP0801).

\vskip 1cm 
\newpage

\begin{center}
{\large\bf Table I}
\end{center}
\vskip 5mm

Quantities for the polytropic stars of different polytropic indices. 

\vskip 5mm
\begin{center}
\begin{tabular}{|c|c|c|c|c|c|c|c|} \hline
$~~n~~$ & $~~~~~~\xi_1~~~~~$ & $~~~~~|\theta'|~~~~~$ & 
$~~~~~k_1~~~~~$ & $~~~~~\kappa_n~~~~~$ & $~~~~~\sigma_n~~~~~$ & 
$~~~~~\lambda_n~~~~~$ & $~~~~~\eta_n~~~~~$ 
 \\ \hline
0 & $\sqrt{6}$ & $2/\sqrt{6}$ & 0 & 1 & $2/35$ & $8/35$ & 6/35 
\\ \hline
0.25&2.59209  & .633717 & .108069 & .903969 & .0561363 & .252922 & .156543
\\ \hline
0.5 & 2.75270 & .499997 & .229390 & .814828 & .0538683 & .273988 & .143283
\\ \hline
0.75& 2.93452 & .398094 & .360772 & .731532 & .0507645 & .292471 & .131209
\\ \hline
1.0 & 3.14159 & .318310 & .500000 & .653455 & .0471121 & .308844 & .120046
\\ \hline
1.5 & 3.65375 & .203301 & .795872 & .511500 & .0389291 & .336503 & .0997846
\\ \hline
\end{tabular}
\end{center}

\newpage
\begin{center}
{\large\bf Figure captions}
\end{center}

\begin{description}
\item[Figs.1] 
The energy($E/(M_*^2/a_*)$, (a)),
the angular momentum($J/\sqrt{M_*^3a_*}$, (b)), and the angular
velocity($\Omega/\sqrt{\pi\rho_*}$, (c)) 
as a function of the orbital separation($R_*/a_*$) for $n=0.5$. 
The solid and dotted lines denote the results calculated by
the ellipsoidal approximation for $C_{\rm s}=0$ and 0.02,
and the open and filled circles denote the results by the numerical
calculation for $C_{\rm s}=0$ and 0.02, respectively.
Note that the innermost point of the sequence is determined 
from the condition $2a_1=R$. 

\item[Figs.2] The same as in figs.1, but for $n=1$ and 
$C_{\rm s}=0$, $1/60$. 

\item[Figs.3] 
The relation between the angular momentum and the
angular velocity for $n=0.5$(a) and 1.0(b). 
The meaning of the lines and marks is the same as that in figs.1 or 2.

\item[Figs.4] 
The angular momentum as a function of $R_*/a_*$ (a) 
and the angular velocity as a function of the angular 
momentum (b) for $n=0.25$, and 
for $C_{\rm s}=0$(solid line), 0.01(dotted line),
0.02(dashed line), 0.03(long dashed line). 

\item[Figs.5] The same as figs.5, but for $n=0.5$. 

\item[Figs.6] The same as figs.5, but for $n=0.75$.

\item[Figs.7] The same as figs.5, but for $n=1.0$.

\item[Figs.8] The same as figs.5, but for $n=1.5$.

\item[Figs.9] $\rho_{\rm c}/\rho_{\rm N}$ as a
function of the orbital separation for $n=0.5$(a), 1.0(b), and 1.5(c). 
In each figure, the solid, dotted, dashed, and long dashed lines
denote the results for $C_{\rm s}=0$, 0.01, 0.02 and 0.03, respectively.

\end{description}


\begin{thebibliography}{99}

\bibitem{TS}
K. Taniguchi and M. Shibata, submitted to Phys. Rev. D. 

\bibitem{LRS}
D. Lai, F. A. Rasio, and S. L. Shapiro, Astrophys. J. Supple. {\bf 88},
205(1993); Astrophys. J. {\bf 420}, 811(1994).

\bibitem{shapiro} For example, S. L. Shapiro and S. A. Teukolsky, {\it 
Black Holes, White Dwarfs, and Neutron Stars}(New York, Wiley, 1983). 

\bibitem{shibapn}
M. Shibata, submitted to Phys. Rev. D; \\
M. Shibata, Prog. Theor. Phys. {\bf 96}, 317(1996).

\bibitem{chand}
S. Chandrasekhar, Astrophys. J. {\bf 142}, 1488(1965).


\bibitem{PNeom}
S. Chandrasekhar, Astrophys. J. {\bf 148}, 621(1967), and also see
nice collection, {\it Selected Papers}, S. Chandrasekhar, Vol.5
(Chicago University Press, 1990).

\bibitem{asada}
H. Asada and M. Shibata,  Phys. Rev. D {\bf 54}, 4944(1996).

\bibitem{Ch69}
S. Chandrasekhar, {\it Ellipsoidal Figures of Equilibrium}
(Yale University Press, New Haven, 1969).

\bibitem{will}
C. M. Will, {\it Theory and Experiment in Gravitational Physics}
(Cambridge University Press, Cambridge, 1981), p.146.

\bibitem{chandra}
S. Chandrasekhar, Astrophys. J. {\bf 140}, 417(1964); {\bf 142}, 1519(1965). 

\bibitem{wiseman}
A. G. Wiseman, gr-qc 9704018.

\bibitem{lai}
D. Lai, Phys. Rev. Lett. {\bf 76}, 4878(1996): He showed that the 
Newtonian tidal effect acts to stabilize each star of binary. 

\bibitem{wilson}
J. R. Wilson, G. J. Mathews, and P. Marrinetti, Phys. Rev. D. {\bf 54},
1317(1996).

\end{thebibliography}
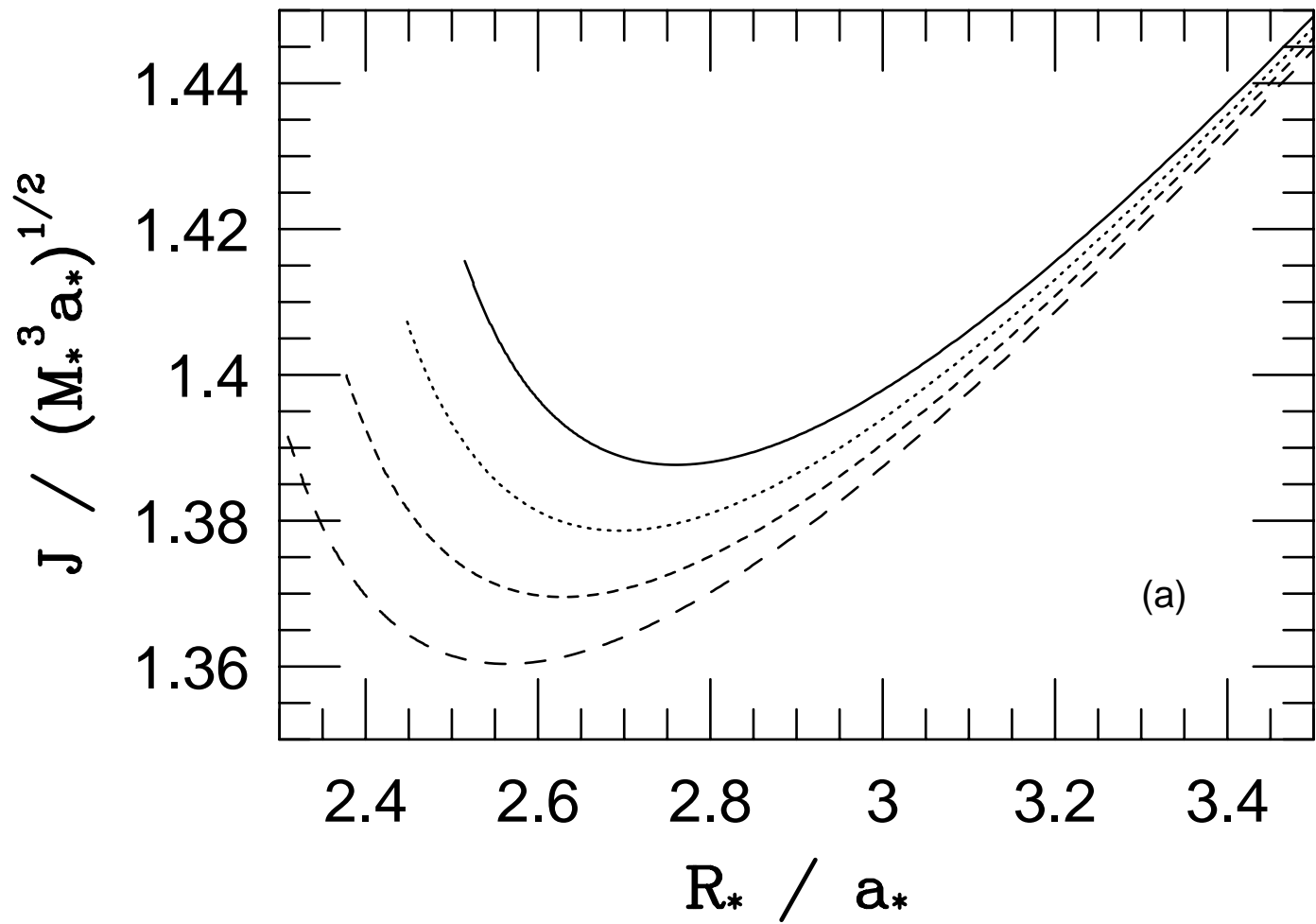
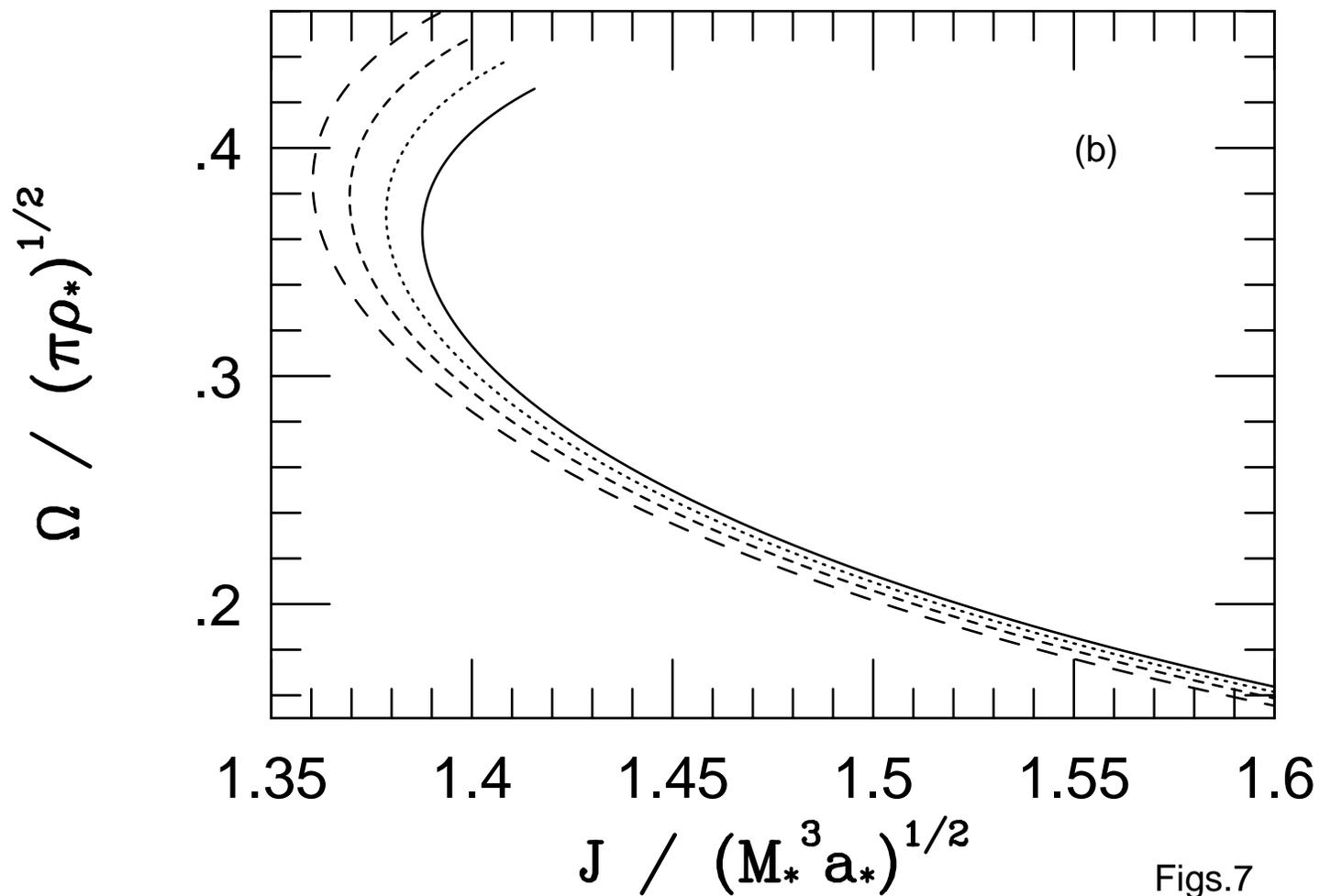
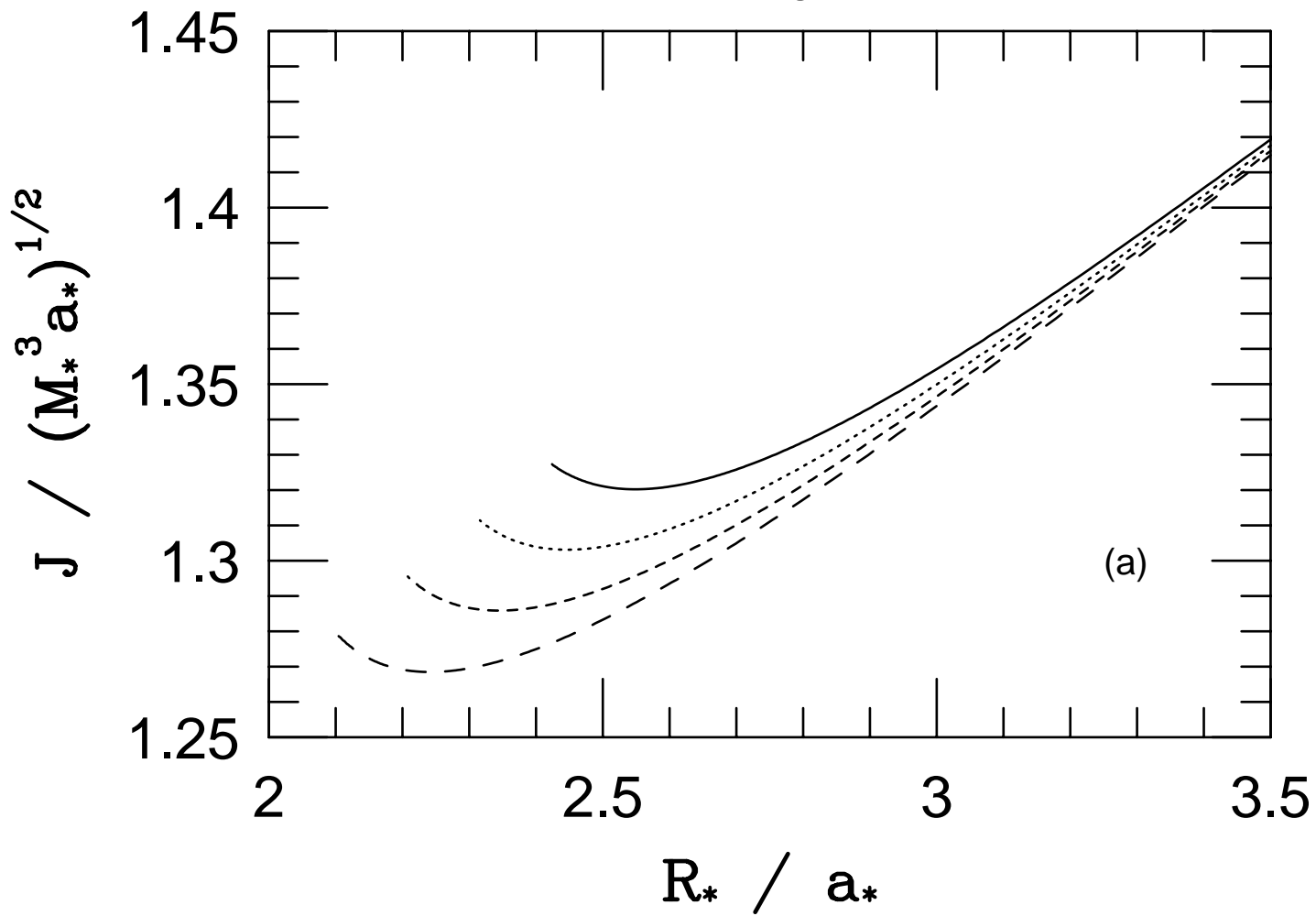
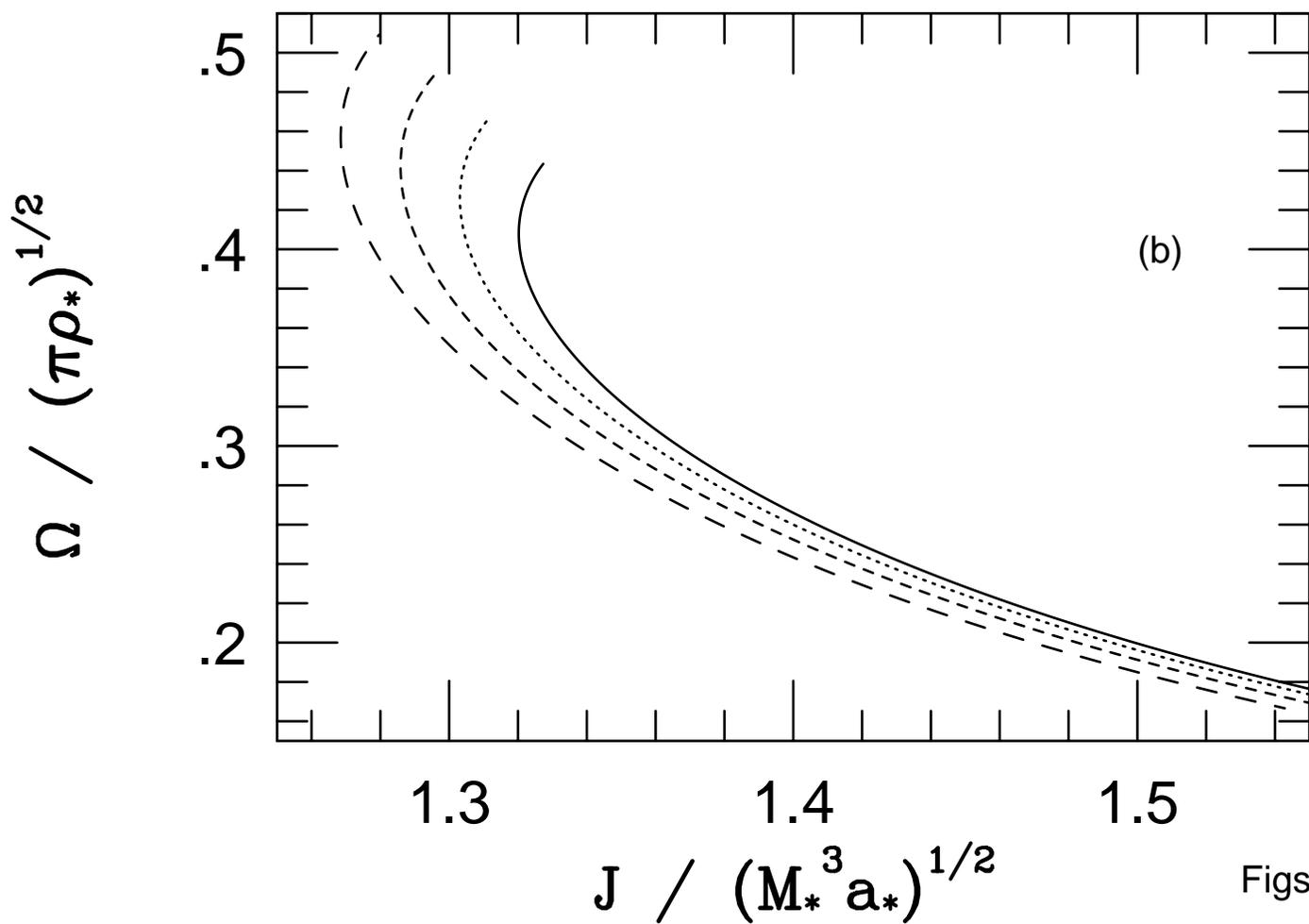
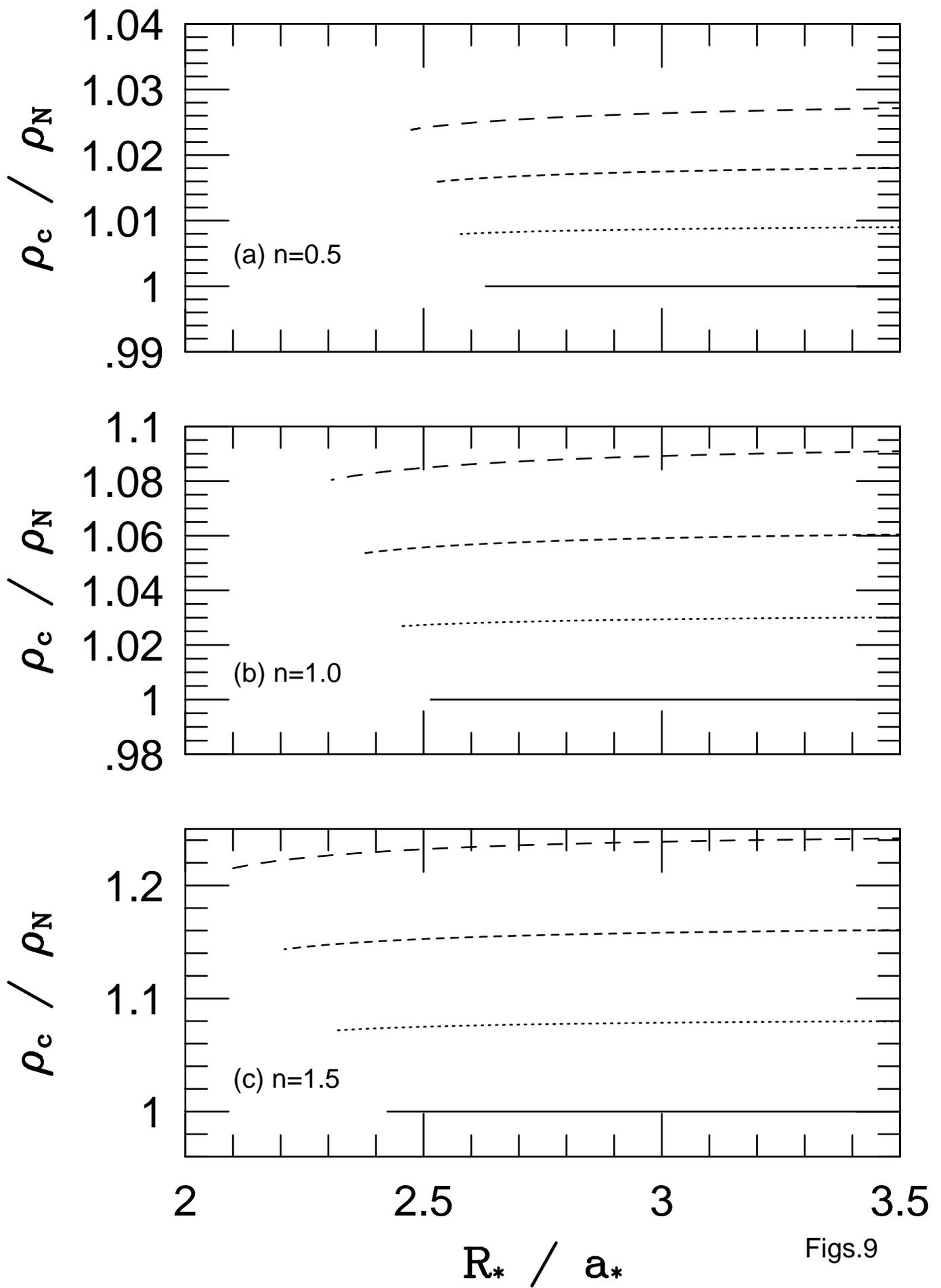
\end{document}